\documentclass[%
 aip,
 pop,%
preprint,%
superscriptaddress
]{revtex4-1}

\usepackage{graphicx}
\usepackage{epsfig} 
\usepackage{wrapfig}
\usepackage{amsmath,amsbsy}
\usepackage{setspace}

\usepackage{natbib}
\setcitestyle{numbers,open={ Ref. }}

\begin{document}

\title{Orbital Spectrum Analysis of Non-Axisymmetric Perturbations of the Guiding-Center Particle Motion in Axisymmetric Equilibria}

\author{P.A. Zestanakis}
\affiliation{ 
School of Electrical and Computer Engineering,  National Technical University of Athens, Greece
}%
 
\author{Y. Kominis}%
\affiliation{ 
School of Applied and Physical Science,  National Technical University of Athens, Greece
}%

\author{G. Anastassiou}
\affiliation{ 
School of Electrical and Computer Engineering,  National Technical University of Athens, Greece
}%

\author{K.Hizanidis}
\affiliation{ 
School of Electrical and Computer Engineering,  National Technical University of Athens, Greece
}%

\date{\today}

\begin{abstract}
The presence of non-axisymmetric perturbations in an axisymmetric magnetic field equilibrium renders the Guiding Center (GC) particle motion non-integrable and may result in particle, energy and momentum redistribution, due to resonance mechanisms. We analyse these perturbations in terms of their spectrum, as observed by the particles in the frame of unperturbed GC motion. We calculate semi-analytically the exact locations and strength of resonant spectral components of multiple perturbations. The presented Orbital Spectrum Analysis (OSA) method is based on an exact Action-Angle transform that fully takes into account Finite Orbit Width (FOW) effects. The method provides insight into the particle dynamics and enables the prediction of the effect of any perturbation to all different types of particles and orbits in a given, analytically or numerically calculated, axisymmetric equilibrium.
\end{abstract}

\maketitle

\section{Introduction}
Guiding Center (GC) theory has been widely used for more than four decades as the basis for the study of single and collective particle dynamics in toroidal magnetic fields utilized in fusion devices\cite{CaryRMP09}. The theory has been originally formulated in a non-canocical\cite{Littlejohn1979} description which was later extended to a canonical one \cite{White_Chance}. The former has the advantage of being applicable to any type of coordinates, whereas the latter, while being more abstract, has the advantage of an elegant description of the underlying dynamics, accompanied by an arsenal of powerful mathematical methods\cite{goldstein1956classical}.

The GC motion of charged particles in an axisymmetric magnetic field is known to be regular, due to the existence of three integrals of motion, namely the energy, the magnetic moment and the canonical toroidal momentum\cite{White_Chance, littlejohn1983}. However, the presence of any non-axisymmetric perturbation results in symmetry breaking, non-integrability and complex particle dynamics. In realistic tokamaks, non-axisymmetric perturbations are introduced due to static magnetic field fluctuations, magnetohydrodynamic (MHD) modes or radio frequency (RF) waves. The effect of these perturbations is the redistribution of particles, energy and momentum, through local (e.g energy absorption) and non-local processes (e.g. energy transport), that are based on resonant interactions with the 3 degrees of freedom of the GC motion\cite{White2010,White2012}.

Significant interaction with any perturbative mode takes place when the mode is resonant with the unperturbed particle motion. Fundamental understanding of the perturbed motion requires the knowledge of the position of the resonant orbits in phase space, where the resonant condition - involving the 3 frequencies of the unperturbed motion - is met. However, this is not sufficient for obtaining a clear picture of the perturbed \textendash single, or collective \textendash particle dynamics. Some measures of the strength and the extent of the resonance in the phase space are also required.

 The collective particle dynamics under symmetry-breaking perturbations can be studied on the basis of single particle GC theory, described above. This is the Gyro-Kinetic (GK) theory, which has been formulated either in non-canonical or canonical coordinates \cite{brizard2007}, with the latter resulting in a kinetic equation of the Fokker-Planck type in the Action space \cite{Kaufman1972}. This is the standard quasilinear transport theory formulated either by the trajectory integral approach \cite{Lamalle199345,VanEester_Koch1998,Brambilla1999} or by the Hamiltonian approach, both made possible due to the canonical structure of the GC phase space \cite{White_Chance,White_Boozer_Hay,Wang2006}. Common to both approaches is the requirement that the quasilinear diffusion operator be expressed in terms of constants of the unperturbed motion. The former approach relies on the application of integration operators on the perturbations, the constants of motion being used to label the unperturbed orbits, whereas in the latter, the constants of motion are the Action variables  \cite{Kaufman1972,Hazeltine1981,Gambier_Samain1985,Kominis2008,KominisPRE2008,Kominis2010}

Although the Action-Angle description is accompanied by powerful mathematical tools \cite{goldstein1956classical,lichtenberg1992regular} and is widely appreciated for its elegance, applications have been restricted to either formal derivations  \cite{Kaufman1972,Gambier_Samain1985,Kominis2008,KominisPRE2008,Kominis2010}  or calculations under strict assumptions\cite{Abdullaev2006,Hazeltine1981}. Explicit calculations of AA variables have been carried out only for the simple case of Large Aspect Ratio (LAR) equilibria for transit and banana orbits, under the approximation of zero drift from the magnetic surfaces or Zero Orbit Width (ZOW) approximation\cite{Hazeltine1981,Brizard2011}.  However, it is known that energetic particle orbits deviate strongly from the magnetic surfaces. Even the simple case of concentric circular magnetic surface equilibria supports 10 orbit types other than the transit and banana orbits\cite{Gott2013,Eriksson2001} of Standard Neoclassical Theory (SNT). 
The effect of such orbits has been long debated \cite{Bergmann2001,Lin1997,Shaing1997,Shaing2004,Helander2000}, and though it seems that the contribution from low energy non\textendash{}standard orbits is not significant\cite{Helander2000}, this cannot be argued for energetic orbits as well, since the bounce and drift frequencies can become comparable\cite{Eriksson2001}, giving rise to new interactions and instabilities, which SNT cannot predict. White and al.\cite{White2010,White2012} have recently provided a method of numerically locating resonances with a particular mode in phase space, by means of the vector rotation criterion. This involves particle tracing, a time consuming process of numerically integrating particle orbits with initial conditions that span the entire phase space for long enough integration times so that the resonance is manifested and for every perturbation separately.

Undoubtedly, the AA formalism provides the appropriate description and concepts for understanding the particle motion, due to its direct relation to the three adiabatic constant of GC dynamics. No wonder the scientific community have decided to include it in most textbooks on plasma physics or fusion (e.g.\cite{chen2013introduction,wessonTokamaks}). The widespread notion that the AA formalism cannot provide specific results for realistic magnetic field configurations (e.g.\cite{VanEester1999,Lamalle199345}) stems from the fact that so far no general method for obtaining the transform from configuration variables to AA variables has been presented, rather than from some intrinsic obscurity of the Hamiltonian formalism itself. So do the aforementioned restrictions and approximations. In fact, we argue that it is the AA formalism that elucidates the dynamics and separates the timescales of  different degrees of motion. After all, it is the AA formalism that takes the most advantage of the Hamiltonian structure and fully exploits the canonical structure of GC dynamics.

In this paper, we demonstrate a method for calculating the transform from configuration space variables to AA variables for any given axisymmetric equilibrium. The existence of a local transform to AA variable is guaranteed by the symmetries of the unperturbed system \cite{goldstein1956classical}, and, though, in general, no such global transform exists, we are able to cover all phase space by calculating multiple AA transforms. The orbital frequencies, being constants of motion, are functions of the Actions alone, so that the resonances can be located and studied on the Action subspace. Since the Actions are both the canonical momenta and the integrals of motion, perturbation analysis is significantly simplified\cite{goldstein1956classical,lichtenberg1992regular}.

Based on the AA transform we introduce the Orbital Spectrum Analysis (OSA) method for analytically estimating the effects of particle interaction with different kinds of perturbations. In OSA, all different kinds of orbits are treated on equal footing, without referring to  phenomenological taxonomies, which makes it straightforward to expand the analysis to equilibria more complex than LAR. One of the most significant advantages this approach has to offer is that the frequencies of the different degrees of freedom are readily calculated and that the resonance condition can be written in a simple form. Full Orbit Width (FOW) effects are intrinsically taken into account and the phase space location as well as the effective strength of resonances is automatically revealed. Moreover, the Actions, being both canonical momenta and constants of motion, provide an excellent framework for building an equilibrium distribution function, a task that until recently has been known to be problematic\cite{DiTroia2012}. 

  In \mbox{section \ref{sec:AAT}}  the canonical GC motion is reviewed and the AA transform algorithm is outlined. In \mbox{section \ref{sec:Chaotic}} we introduce the Orbital Spectrum Analysis (OSA) method and demonstrate it by applying it to the case of synergetic interaction with two nonresonant magnetic perturbations, where chaotic particle motion occurs, while the magnetic field lines remain regular and no magnetic surface destruction occurs. By means of OSA, the conditions for transition to chaos are analytically determined.

\section{The Action Angle Transform}
\label{sec:AAT}
 
The Lagrangian of the Guiding Center (GC) motion of a charged particle is $\mathcal{L} = \left(\mathbf{A}+\rho_\parallel\cdot \mathbf{B}\right)\cdot\mathbf{v} + \mu \dot{\xi} - H$, where $\mathbf{A} $ and $\mathbf{B}$ are the vector potential and the magnetic field respectively, $\mathbf{v}$ is the guiding center velocity, $\mu$ the magnetic moment, $\xi$, the gyrophase, $\rho_\parallel$ the parallel velocity to the magnetic field, normalized with $B$ and 
\begin{equation}
H =\rho_\parallel^2 B^2/2 + \mu B +\Phi
\label{eq:Hamiltonian_conf}
\end{equation}
the Hamiltonian, with $\Phi$ the electric potential \cite{littlejohn1983} (see Appendix). All quantities are evaluated at the guiding center position and normalized with respect to the nominal magnetic axis gyrofrequency and the major radius $R$. It has been shown that, when the magnetic coordinates $(\psi,\tau,\chi)$ \textendash{} $\psi$ being the toroidal flux, $\tau$ and $\chi$ the poloidal and toroidal angle \textendash{} are appropriately chosen, the dynamical system is Hamiltonian and one can define $P_{\tau}$ and $P_{\chi}$ to be the canonical poloidal and toroidal momenta respectively \cite{White_Chance}. In axisymmetric equilibria, the canonical position $\chi$ is ignorable and $P_\chi$ is conserved, so that the dynamical system, being reduced to one Degree Of Freedom (DOF), is integrable. However, the motion in phase space is non\textendash{}trivial and there is no straightforward way to predict the behaviour of the system when perturbations are introduced and the integrability is lost.

\begin{figure}[htb!]
    \centering
        \centering
        \includegraphics[scale=1]{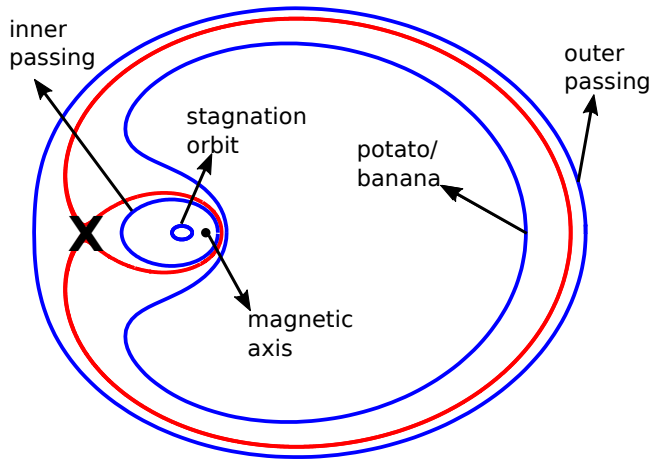}
        \caption{\linespread{1} \fontsize{10pt}{12pt}\selectfont Schematic diagram of LAR characteristic orbits. Two separatrices, homoclinic to the x-point, are shown, each one acting as a boundary between two continents.}
        \label{fig:characteristic orbits}
\end{figure}

The conserved canonical momenta $P_\chi$ and $\mu$ are already the actions of the toroidal motion and the gyromotion respectively. The AA pair ($J, \theta$) of the poloidal motion is found by integrating along a closed orbit in the poloidal plane:
\begin{equation}
J\ =\ \frac{1}{2 \pi} \oint P_\tau d\tau
\label{eq:J}
\end{equation}
and defining $\theta$ as the normalized time
\begin{equation}
\theta\ =\ \omega_\theta \left(J,P_\chi,\mu\right) t,
\label{eq:theta}
\end{equation}
where $\omega_\theta$ is the frequency of the poloidal motion and depends only on the three actions.
By virtue of the Liouville\textendash{}Arnlold theorem\cite{arnold1989mathematical}, such a transform always exists locally. In particular we can cover all phase space with a measurable set of AA transforms, one for each phase space region that is bounded by a separatrix. From now on we shall call such a region a \textit{continent}, while the set of transforms for all regions is called an \textit{atlas}.  

\mbox{Fig. \ref{fig:characteristic orbits}} depicts a poloidal projection of LAR orbits with the same $P_\chi$ and $\mu$. Each of the two separatrices (thick red lines) separates two continents. All orbits belonging to the same continent share the same topology and the behaviour of the dynamics varies continuously within each continent. The conventional orbit labelling (e.g. passing, trapped, stagnated) is phenomenological and refers to the magnetic axis, because it is the natural reference point of the magnetic field geometry and very close to the natural reference point of the slow particle dynamics, i.e. the elliptic point of the inner passing orbits (first continent in the figure). However, for the energetic particle dynamics, the magnetic axis is not special from the dynamics point of view, and conventional labelling is arbitrary and confusing.
For example, in \mbox{Fig. \ref{fig:characteristic orbits}}, the two orbits marked as stagnation and inner passing orbits share the shame topology, but their conventional labelling is not the same.
In fact, it would be more useful to label the orbits with respect to the continent they belong to. This choice not only underlines the dynamic characteristics of each orbit, but serves  the purpose of clarity and algorithmic simplicity.

Although the Hamiltonian $H$ is integrable, finding an analytic solution is impractical. Instead, we calculate the atlas numerically. The algorithm we follow relies on finding the boundaries of each continent in $(\tau,\psi,P_\chi,\mu)$  subspace, i.e. the separatrices,
and then calculating numerically each AA transform defined by \mbox{eqs. \ref{eq:J}, \ref{eq:theta}} on a carefully chosen sample of closed orbits. The particulars of the algorithm will be discussed in another paper.

In transforming from $(\tau,\psi)$ 
 to $(J, \theta)$, the angles of the other two DOFs are redefined, so that eventually all DOFs are described by AA pairs. In each continent, the transform is comprised from a non\textendash canonical step, from $(\tau,\psi,P_\chi,\mu)$ to $(\tau,P_\tau,P_\chi,\mu)$ (see Appendix), and a canonical one to $\left(\theta, J, P_\chi, \mu\right)$, which is implicitly generated by a function of the form $F_2\ =\ F\left(\tau, J, P_\chi, \mu\right)$. Dependence of $F_2$ on  $P_\chi$ and $\mu$ implies that $\chi$ is also transformed to an angle variable 
\begin{equation}
\hat \chi\ =\ \chi - f_\chi\left(J ,\theta,  P_\chi, \mu\right)
\label{eq:chi_bar}
\end{equation}
and so does $\xi$. Therefore, this procedure generates the transform to the AA pairs 
$\left(P_\chi, \hat \chi\right)$ and $\left(\mu, \bar \xi\right)$. 
The barred angles differ from the original in that they evolve linearly with time,  with frequencies $\omega_{\hat \chi}$ and $\Omega_c$, which are the averaged drift frequency and the mean gyrofrequency respectively. By construction, the new poloidal angle $\theta$ evolves linearly as well, with the average poloidal frequency $\omega_\theta$, while  the actions $J$, $P_\chi$ and $\mu$ are constants of motion.

\section{Chaotic Motion, Due To Nonresonant Magnetic Fields}
\label{sec:Chaotic}
We will demonstrate the power of the AA transform, by applying it on the study of the particle dynamics in the presence of nonresonant magnetic perturbations. As we show below, the particle motion can become chaotic, even though there is no destruction of the magnetic surfaces. This should come as no surprise, because, only only under the ZOW approximation is the chaoticity of the particle motion directly related to the chaoticity of the magnetic field lines. In this case, however, the drift effects are significant, so that the stochastisation of magnetic field lines is not a reliable criterion for predicting chaotic particle motion, when FOW effects are involved. Resonance and resonance overlap is a dynamic effect, the analysis of which, ideally, should be restricted on the particulars of phase space alone, without resorting to configuration geometry concepts. 
This is, of course, not the first time such a behaviour has been described (see, for example, \cite{Matsuyama2014} and references therein), but here it is analysed in the action space alone, without referring to the configuration geometry. The simplicity of this approach and the excellent quantitative results it can provide is a major advantage of the OSA method.

\subsection{Orbital Spectrum Analysis}
\label{subsec:OSA}
A perturbation of the form $\delta\mathbf{B} = \pmb{\nabla}\times \sigma \mathbf{B}$ can be straightforwardly included in the guiding center Hamiltonian as $H =(\rho_c -\sigma)^2 B^2/2 + \mu B +\Phi$, $\rho_c = \rho_\parallel + \sigma$. \cite{white2001theory} This introduces a first and second order perturbation Hamiltonian term. Any perturbation with physical meaning should be given in terms of the magnetic coordinates, or even the lab coordinates. The first order Hamiltonian $H_1$ is proportional to $\sigma$. Due to the nonlinear depencence of $\hat \chi$ on $\theta$ (\mbox{eq. \ref{eq:chi_bar}}), a monochromatic mode $\sigma =  A_{m,n}\left(\psi\right) \mathrm{exp}\left(i\left(m\chi +  n\tau -\omega t\right)\right)$ in the magnetic coordinates gives an infinite series of modes in the AA phase space, so that 
\begin{equation}
H_1 = \sum_{s}{\mathcal{H}^1_{s,m}}\left(J,P_\chi\right)\ e^{i\left(m\hat\chi +  s\theta - \omega t\right)},
\label{eq:AA_p_s_perturbation}
\end{equation}
where 
\begin{equation}
\mathcal{H}^1_{s,m}\left(J,P_\chi\right) = \frac{1}{2\pi} \oint\limits_{J,P_\chi=\mathrm{const.}} H^1_{m,n}\left(\psi\right)
 \ e^{i\left(n\tau + n f_\chi\left(J,P_\chi,\theta\right) -s\theta\right)} d\theta.
\label{eq:AA_p_s_spectral_transform}
\end{equation}

As equation \mbox{eq. \ref{eq:AA_p_s_perturbation}} indicates, the resonances of the perturbation  are located in action space at the points where the resonance condition
\begin{equation}
m\ \omega_{\hat \chi}\left(J,P_\chi,\mu \right) +  s\ \omega_\theta\left(J,P_\chi,\mu \right) - \omega\  =0
\label{eq:resonance_condition}
\end{equation}
is met. The location of the resonances in action space depends only on the spectral parameters $m$ and $\omega$. The actual profile of the perturbation, i.e the depencence on $n$ or $\psi$, is relevant in defining the amplitude of the resonant terms, but not in pinpointing their location in the orbital spectrum. Since $s$ can take on any integer value, each bounded continent contains an infinite number of such frequencies, most of which are located in the narrow chaotic sea near the separatrix, where $\omega_\theta$ approaches zero. In the bulk of each continent there are only a few, if any, sites where \mbox{eq. \ref{eq:resonance_condition}} is satisfied. 

 Near a particular resonance $m\ \omega_{\hat \chi}+  s\ \omega_\theta -\omega = 0$, the dynamics follow a pendulum\textendash{} like Hamiltonian and a trapped area of width proportional to the square root of the perturbation amplitude is formed. The width depends on $\mathcal{H}_{s,m}\left(J,P_\chi\right)$ and can be easily calculated once the AA transform has been performed.
 
 \begin{figure}[htb!]
        \includegraphics[scale=1]{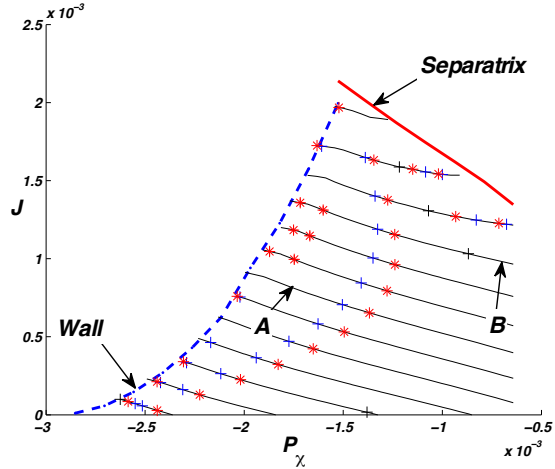}
        \caption{Resonance chart cross section in $\mu$. The solid black lines depict the energy surfaces, crosses and stars correspond to resonances with $m=10$ and $m=8$ respectively.}
        \label{fig:Resonance chart}
\end{figure}
 
\subsection{Application}

We apply this analysis on the study of the dynamics in a peaked equilibrium (see Appendix) in the presence of two static magnetic perturbations with $m=8$ and $m=10$. The safety factor ranges from 1 to 1.8, so that the perturbations are nonresonant and no flux surface is destroyed. The amplitudes of the perturbation are of the order $\delta B/B \approx 10^{-4} - 10^{-3}$, well inside the domain of validity of perturbation theory.

This case of time independent perturbations is particularly simple, but rather indicative of the power of the AA transform and the OSA method. The Hamiltonian is conserved and the quantity $P_\chi - m/s J $ is an adiabatic invariant. The cases where the ratio $m/s$ or $s/m$ becomes very large are of little interest, since the adiabatic invariant coincides with one of the actions, so that no significant redistribution takes place.

Without any particular knowledge about the actual profile of the perturbations, other than the toroidal numbers, it is possible to chart the location of the resonances in the action space of each continent, by requiring that the condition
\begin{equation}
\frac{m \omega_{\hat \chi}}{\omega_\theta}\ =\ \textrm{integer}.
\end{equation}
\mbox{Fig. \ref{fig:Resonance chart}} depicts the chart of the $m=10$ and $m=8$ resonances in a potato\textendash{}banana continent. The constant energy subsurfaces, on which the perturbed motion will be confined, due to time independence of the perturbation, are plotted with thick black lines.

\begin{figure}[htb!]
         \includegraphics[scale=1]{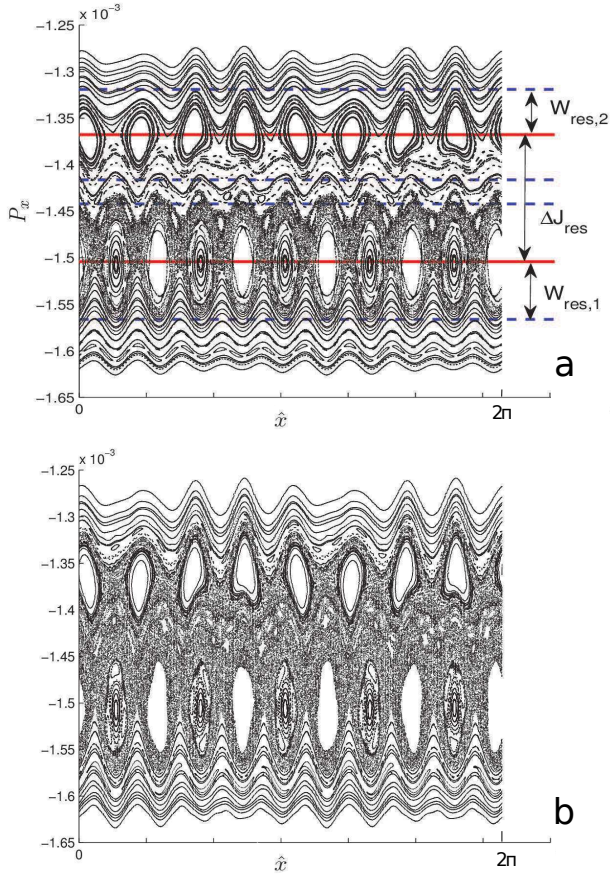}
    \caption{ Inspection of the resonance chart can reveal the phase space regions when mode synergy can be significant. The analytically calculated resonance positions, width and overlap conditions are in excellent agreement with the simulations. a) Poincare plot on the surface A of \mbox{Fig. \ref{fig:Resonance chart} } for two modes with subcritical amplitude. The semianalitically calculated positions of the resonances as well as their widths are denoted with solid and dashed lines respectively. b) The same Poincare plot for perturbations with critical amplitude. KAM lines between the two resonances have been destroyed and significant redistribution can take place.}
    \label{fig:resonance interplay}
\end{figure}

Coexistence of more than one resonances on the same energy surface can lead to chaotic redistribution, due to destruction of the adiabatic invariant. On the energy surface denoted by A in \mbox{Fig. \ref{fig:Resonance chart}} there are two neighbouring resonances located at \mbox{$P_{\chi,1} \approx -1.5\cdot 10^{-3}$} and\mbox{ $P_{\chi,2} \approx -1.37\cdot 10^{-3}$}.  Chiricov criterion\cite{lichtenberg1992regular} provides a simple estimation for the onset of chaotic motion, by requiring that the resonances overlap.  The resonance widths $W_{\textrm{res},i}$ are estimated by approximating the motion around the resonances with the pendulum Hamiltonian and using the well known formula for the separatrix width. The criterion requires that for chaotic motion
\begin{equation}
\Delta J \leq W_{\textrm{res},1} + W_{\textrm{res},2},
\end{equation}
where $\Delta J$ is the distance between two neighbouring resonances.

As demonstrated in \mbox{Fig. \ref{fig:resonance interplay}}, the OSA method predicts both the location of the resonance center as well as the resonance width in the phase space. Moreover, application of the Chirikov criterion is shown very successful in predicting the transition from weak to strong chaos. \mbox{Fig. \ref{fig:resonance interplay}a} displays a Poincare plot of the perturbed motion, when the amplitude of the perturbation $A = A_\textrm{subcrit}$ is lower than the analytically computed critical amplitude $A_\textrm{crit}$. Some chaos is present, due to the existence of higher order resonances, but the primary resonances are well separated by KAM surfaces and no significant particle redistribution takes place. Superimposed on the Poincare plot are the analytically calculated locations of the resonances (solid lines) and the resonance widths (dashed lines). It is evident that the resonances do not overlap. The situation changes for $A = A_\textrm{crit} = A_\textrm{Chiricov}$ (\mbox{Fig. \ref{fig:resonance interplay}b}), for which the KAM surfaces have been destroyed and there is a chaotic sea between the two primary resonances. Some stickiness that can be observed a little lower than the upper chain of primary islands is reminiscent of the last KAM surface to be destroyed.

\begin{figure}[htb!]
    \centering
        \includegraphics[scale=1]{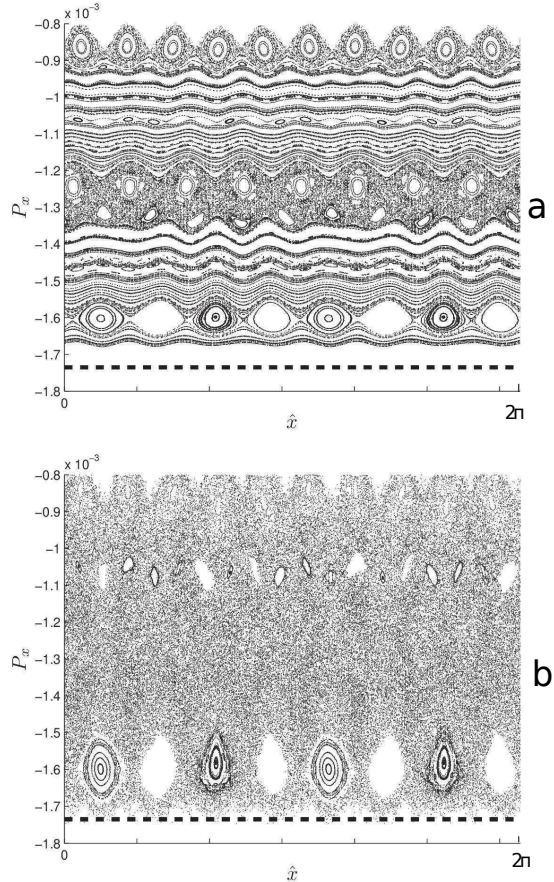}
   \caption{The OSA method as a tool for estimating conditions for confinement loss. The outer closed flux surface is marked with a thick dashed line. a) Poincare cut for the energy surface B of  \mbox{Fig. \ref{fig:Resonance chart}} and subcritical amplitude $0.08\ A_{\mathrm{Chirikov}}$. Only two of the resonances have partially overlapped. b) The same, with amplitude $0.3\  A_{\mathrm{Chirikov}}$. Although, this is still below the critical value determined by Chirikov criterion, the KAM surfaces have been destroyed. Chirikov criterion overestimates the critical amplitude, by ignoring higher order resonances.}
    \label{fig:from_confinement2loss}
\end{figure}

The case of the energy surface B in \mbox{Fig. \ref{fig:Resonance chart}} is of particular importance, because there is a line of resonances linking a deeply trapped part of phase space to the plasma wall (dashed blue line).  Therefore, provided that all consecutive resonances overlap, significant particle loss can occur. Chirikov criterion determines the critical amplitude at $A_\textrm{Chirikov} = 1.2\cdot 10^{-3}$,which is an overestimation of the critical amplitude $A_\textrm{crit}$, due to the strong presence of higher order resonances, as demonstrated in \mbox{Fig. \ref{fig:from_confinement2loss}}.

\begin{figure}[htb!]
    \centering
        \includegraphics[scale=1]{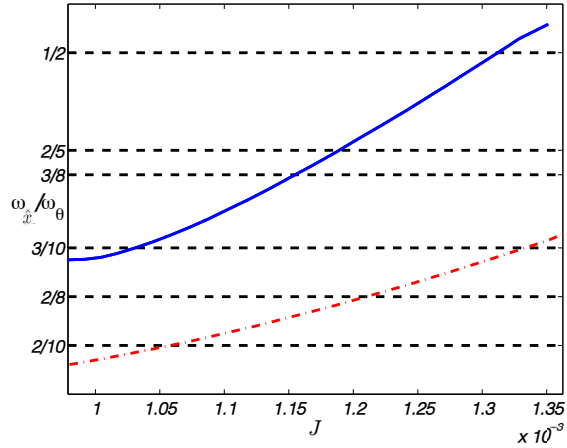}
        \caption{Toroidal over poloidal frequency ratio as a function of $J$ on the energy surface B. The resonances with the $m=8$ and $m=10$ are located at the intersections with the horizontal dashed lines. Solid curve: numerical calculation with full orbit width effects. Dashed-dotted curve: analytic zero orbit width approximation.}
        \label{fig:Bresonances}
\end{figure}

It is important to point out that zero orbit width approximations, which are necessary in order to calculate an analytical AA transform, cannot reproduce the picture described above. 
The location of the resonances is determined by requiring that $\omega_{\hat \chi}/\omega_\theta = s/10$ or $\omega_{\hat \chi}/\omega_\theta = s/8$, where $s$ is any integer. \mbox{Figure \ref{fig:Bresonances}} depicts the numerically calculated FOW frequency ratio curve, as well as the analytical one, calculated using well known formulas for ZOW approximation \cite{white2001theory,Brizard2011}. 
As is clear from the figure, the two approaches predict entirely different resonance ratios and locations. ZOW approximation, therefore, fails for energetic particles, even in the case of a LAR equilibrium, where it is expected to be more accurate. 

The previous analysis refers to the phase space study of the single particle motion. The resulting kinetic modelling of the collective dynamical behaviour consists of a Focker\textendash{}Plank equation for the evolution of the Angle-averaged distribution function in the Action space, when the transition to strong chaos has occurred\cite{lichtenberg1992regular,Kaufman1972Reformulation}. The three Action variables are related to magnetic moment, parallel momentum and radial position (or energy) so that particle, momentum and energy transport is described by the corresponding Action-dependent quasilinear diffusion tensor.

\begin{figure}[htb!]
    \centering
        \includegraphics[scale=1]{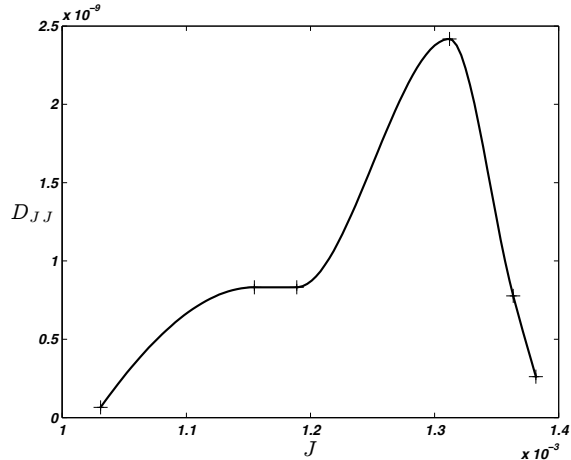}
        \caption{The $J,J$ element of the quasilinear tensor for the case of \mbox{Fig. \ref{fig:from_confinement2loss}b}.}
        \label{fig:quasilinear}
\end{figure}

Assuming that the gyromotion is much faster than every other process involved, the magnetic moment remains constant and the resonance condition \mbox{Eq. \ref{eq:resonance_condition}}, which does not involve the gyrofrequency $\Omega_c$, applies \cite{Kaufman1972}. At the points where the resonance condition is met, the diffusion tensor is given by
\begin{equation}
\pmb{\bar{D}} = \pi\sum\limits_\mathbf{r} \mathbf{r}\mathbf{r}
\left|\mathcal{H}^1_\mathbf{r}\right|^2\delta(\mathbf{r}\cdot\pmb{\omega}),
\label{eq:quasilinear_diffusion_tensor}
\end{equation}
and the Focker\textendash{}Plank equation takes the form
\begin{equation}
\frac{\partial}{\partial t}f\left(\pmb{I},\mu;t\right)\ =\ \frac{\partial}{\partial \pmb{I}}\cdot\left(\pmb{D}\cdot \frac{\partial}{\partial \pmb{I}} f\left(\pmb{I},\mu;t\right)\right)
\end{equation}
where $\mathbf{r} = (s,m)$ is the vector of the harmonic numbers, $\pmb{\omega} = (\omega_\theta, \omega_{\hat \chi})$, $\pmb{I} = (J,P_\chi)$ and $f$ is the distribution function  \cite{Kaufman1972Reformulation, abdullaev2006book}. The diffusion tensor $\pmb{D}$ used in the Focker\textendash{}Plank equation is a smooth function that interpolates the singular diffusion tensor $\pmb{\bar{D}}$ at the resonance points. This is because of the spreading of resonances, due to nonlinear effects \cite{Kaufman1972}. The fact that the diffusion tensor can be expressed explicitly in Action space is one of the many advantages of the AA formalism. \mbox{Fig. \ref{fig:quasilinear}} depicts the $J,J$ element of the diffusion tensor as a function of $J$ for the case of \mbox{Fig. \ref{fig:from_confinement2loss}b}.

\section{Summary and Conclusions}
The Orbital Spectrum Analysis (OSA) method of any type of non-axisymmetric pertirbation on the GC motion has been presented. The method is based on the Action-Angle (AA) transform of the unperturbed motion allowing for the calculation of the orbital frequencies of the three degrees of freedom and the explicit expression of the resonant conditions as well as of the perturbations in AA variables. It is shown that the exact location and the strength (width) of the resonances in the GC phase space can be calculated, providing an apriori knowledge (without tracing particles) of the effect perturbations including synegetic effects between different types of perturbations. It is worthmentioning that the AA transform is determined from each specific axisymmetric equilibrium and completely characterizes it in terms of the GC motion. Therefore, it has to be calculated only once for a given equilibrium and then it is available for the orbital spectrum analysis of any perturbation. The OSA method serves as an ideal tool for phase space engineering of the synergetic of antagonistic effects of different types of perturbations.   

\begin{acknowledgements}
Supported in part by the National Programme on Controlled Thermonuclear Fusion associated with the EUROfusion Consortium in the framework of EJP Co-fund Action, Number SEP-210130335. 
\end{acknowledgements}

\appendix*
\section{Flux coordinates}
\label{sec:Flux}
The White\textendash{}Boozer notation is summarised\cite{white2001theory}. 
In a nested flux surface magnetic field configuration  the magnetic field can be represented as
\begin{equation}
\pmb{B}\ = \pmb{\nabla}\psi \times \pmb{\nabla}\theta
+ \pmb{\nabla}\zeta \times \pmb{\nabla}\psi_p,
\end{equation}
where $\theta $ is the poloidal angle, $\psi$ is the toroidal flux variable, $\psi_p(\psi)$ is the poloidal flux variable. This is the \textit{contravariant} representation of the field. The \textit{covariant} representation is given by
\begin{equation}
\pmb{B}\ = \hat{\delta}\left(\psi,\theta,\zeta\right) \pmb{\nabla}\psi + 
\hat{I}\left(\psi,\theta,\zeta\right)  \pmb{\nabla}\theta
+ \hat{g}\left(\psi,\theta,\zeta\right)\pmb{\nabla}\zeta. 
\end{equation}

There is always a transform of the angular variables $\theta,\zeta$ to the variables $\tau,\chi$ so that the Jacobian $\mathcal{J}$ of the new system $(\psi,\tau,\chi)$ satisfies $\mathcal{J} = f(\psi)/B^2$. This is the Boozer representation\cite{Boozer1980}. The covariant representation then becomes
\begin{equation}
\pmb{B}\ = {\delta}\left(\psi,\theta,\zeta\right) \pmb{\nabla}\psi + 
{I}\left(\psi\right)  \pmb{\nabla}\tau
+ {g}\left(\psi\right)\pmb{\nabla}\chi. 
\end{equation}
In this representation, the equations of motion are canonical. The  canonical momenta conjugate to $\tau, \chi$  are 
\begin{equation}
P_\tau\equiv\psi +\rho_\parallel I
\end{equation}
and 
\begin{equation}
P_\chi\equiv \rho_\parallel g +\psi_p
\end{equation}
respectively.  The Hamiltonian in \mbox{eq. \ref{eq:Hamiltonian_conf}}  takes the form
\begin{equation}
H = \frac{\left(P_\chi +\psi_p\left( P_\tau,P_\chi\right)\right)^2}{2 g^2\left( P_\tau,P_\chi\right)} B^2 + \mu B + \Phi,
\end{equation}
 When $B$ and $\Phi$ are independent of the generalized toroidal angle $\chi$ (quasisymmetric configuration), $P_\chi$ is a constant of motion.

The safety factor, defined as $q=\pmb{B}\cdot\pmb{\nabla}\chi/\pmb{B}\cdot\pmb{\nabla}\tau$, is determined by the balance of the pressure and the magnetic force and, as its name suggests, its profile is an important characteristic of the equilibrium. It is a flux function in Boozer coordinates. It is known that, for LAR equilibria, q profiles of the form
\begin{equation}
q=q_0\left(1+\left(\psi/\psi_0\right)^\nu\right)^\frac{1}{\nu}
\end{equation}
are acceptable solutions of the force balance condition. Equilibria with $\nu = 1,2,3$ are referred to as peaked, rounded and flat respectively \cite{white2001theory}.

\end{document}